\documentclass[aps,pre, reprint, showpacs,superscriptaddress]{revtex4-1}

\usepackage[english]{babel}
\usepackage{amsmath}
\usepackage{graphicx}
\usepackage{setspace}
\usepackage{bm}
\usepackage{color}
\usepackage{amssymb}
\usepackage{dsfont}
\usepackage{hyperref}
\definecolor{hreflinkcolor}{rgb}{0.13,0.17,0.83}
\hypersetup{colorlinks=true,urlcolor=hreflinkcolor,
        linkcolor=hreflinkcolor,citecolor=hreflinkcolor}

\newcommand{\refref} [1] {Ref.~\cite{#1}}

\newcommand{\refeq}  [1] {Eq.~(\ref{#1})}
\newcommand{\refneq}  [1] {(\ref{#1})}
\newcommand{\refEq}  [1] {Equation~(\ref{#1})}
\newcommand{\refeqs} [2]{Eqs.~(\ref{#1})--(\ref{#2})}
\newcommand{\refEqs} [2]{Equations~(\ref{#1})--(\ref{#2})}

\newcommand{\reffig} [1] {Fig.~\ref{#1}}

\newcommand{\refsect}[1] {Sec.~\ref{#1}}

\begin{document}
\singlespace
\title{Relativistic quasi-solitons and embedded solitons with circular polarization in cold plasmas}

\author{G. S\'anchez-Arriaga}
\affiliation{Bioengineering and Aerospace Engineering Department,
Universidad Carlos III de Madrid, Legan\'es, Spain}

\author{E. Siminos}
\affiliation{Department of Physics, Chalmers University of Technology, Gothenburg, Sweden}

\begin{abstract}
The existence of localized electromagnetic structures is discussed
in the framework of the 1-dimensional relativistic Maxwell-fluid
model for a cold plasma with immobile ions. New partially localized
solutions are found with a finite-difference algorithm designed to
locate numerically exact solutions of the Maxwell-fluid system.
These solutions are called quasi-solitons and consist of a localized
electromagnetic wave trapped in a spatially extended electron plasma
wave. They are organized in families characterized by the number of
nodes $p$ of the vector potential and exist in a continuous range of
parameters in the $\omega\--V$ plane, where $V$ is the velocity of
propagation and $\omega$ is the vector potential angular frequency.
A parametric study shows that the familiar fully localized
relativistic solitons are special members of the families of
partially localized quasi-solitons. Soliton solution branches with
$p>1$ are therefore parametrically embedded in the continuum of
quasi-solitons. On the other hand, geometric arguments and numerical
simulations indicate that $p=0$ solitons exist only in the limit of
either small amplitude or vanishing velocity. Direct numerical
simulations of the Maxwell-fluid model indicate that the $p>0$
quasi-solitons (and embedded solitons) are unstable and lead to wake
excitation, while $p=0$ quasi-solitons appear stable. This helps
explain the ubiquitous observation of structures that resemble $p=0$
solitons in numerical simulations of laser-plasma interaction.
\end{abstract}


\maketitle

\section{Introduction}

Relativistic solitary waves are localized structures consisting of a light
wave trapped in a self-generated plasma cavity.
They have received considerable attention over the past decade
both due to their theoretical interest and their experimental relevance.
Relativistic solitary waves have been found as exact solutions
of the Maxwell-fluid model both in 1-dimension,
with circular~\cite{Kozlov_79,Kaw_92,Esirkepov_1998,Farina01a,Sanchez_2011a,Sanchez_2011b},
and linear~\cite{Sanchez_2015} polarization as well as in 2-dimensions~\cite{Sanchez_11}.
As observed in numerical simulations
\cite{Bulanov_92,Bulanov_99,Esirkepov_02,Wu_2013} and experiments
\cite{Borghesi_02,Chen_07,Pirozhkov_07,Sarri_10,Sylla_12}, these
structures are easily excited during the interaction of high-intensity
lasers with plasmas. Although the underlying mathematical model is
non-integrable, they are normally referred as  solitons instead of
solitary waves, a term that will be used in this work.

This work studies  1-dimensional circularly polarized solitons,
which are probably the solutions that received the greatest
attention in the past. In Ref.~\cite{Esirkepov_02} an analytical
solution for standing solitons was found in the case of immobile
ions. It was shown that this solution exists within the continuous
range $\sqrt{2/3}<\omega<1$, where $\omega$ is the normalized
frequency of the vector potential (Sect.~\ref{Sec:Model} discusses
our normalizations). Ref.~\cite{Farina01a}  showed that solitary
waves with velocity $V$ are organized in branches in the $\omega-V$
plane for both mobile and immobile ions. Each branch corresponds to
a solution with a different number $p$ of zeros or nodes of the
vector potential. A large number of branches were computed in Refs
.~\cite{Poornakala_02,Bai_06}. The maximum amplitude of the
waves, where the  branches end, correspond to nonlinear wave
breaking. This mechanism  has been proposed to generate fast ions
\cite{Farina01a}. Ion (electron) dynamics is behind the wave
breaking of the low-node-number (high-node-number) waves.

The continuous spectrum of single-humped ($p=0$) standing waves
($V=0$)~\cite{Esirkepov_02} and the discrete spectrum of  moving
($V\neq 0$) waves~\cite{Farina01a} led to a natural question: does a
smooth transition exist between both cases? The results of
Ref.~\cite{Farina01a} indicated a negative answer, since no moving
solitons with $p=0$ were identified. On the other hand, in
Ref.~\cite{Poornakala_02}, the authors gave a positive answer on the
basis of numerical studies and concluded that a continuous spectrum
exists for moving solitary waves with $p=0$. In other words, it was
claimed that, for a given value of the velocity, finite amplitude
solitons moving with finite velocity exist within a certain
frequency range. In the limit $V\rightarrow 0$ this family would be
a continuation of the $V=0$ solitons (which exist for
$\sqrt{2/3}<\omega<1$), while in the small amplitude limit it would
be a continuation of nonlinear Schr\"odinger equation (NLS)
solitons. While later works~\cite{Farina05,Saxena_06,Saxena_07}
accepted the existence of a continuous spectrum of  $p=0$ solitons,
we show in \refsect{Sec:Solitary:Waves} that such a spectrum would
contradict general geometric arguments from the theory of reversible
dynamical systems~\cite{champneys1998,Mielke_92}. Through a detailed
numerical study we show that a continuous spectrum of moving $p=0$
solitons exists only in one of the two integrable limits of the cold
fluid model: (1) of small amplitude solutions, (2) of vanishing
velocity $V\rightarrow0$.

A further question then arises: how are we to reconcile the
ubiquitous observation of finite-amplitude structures that resemble
$p=0$ moving solitons in numerical simulations with the fact that
such solutions only exist in the small amplitude limit? As we show
in \refsect{Sec:Quasisoliton}, a more general class of partially
localized solutions of the Maxwell-fluid system exists. The
electromagnetic field in these solutions is localized; however, the
plasma density depression exhibits non-vanishing oscillations at its
tail. By analogy to similar solutions that have been observed in
nonlinear optics, we refer to such solutions as
\emph{quasi-solitons}. The familiar $p\neq0$ soliton solution
branches are parametrically embedded inside the continuous spectrum
of quasi-soliton solutions. Moreover, our numerical simulations
indicate that $p=0$ quasi-solitons are stable; this helps shed light
to the abundance of such structures in laser-plasma interaction
simulations.

This paper is organized as follows. Section~\ref{Sec:Model}
introduces the Maxwell-fluid model and the dynamical
system associated with solitons with fixed ions. The properties of this dynamical system are summarized
with emphasis at its conservative and reversible
character. These properties are used in Sec.~\ref{Sec:Geometric:Arguments}
to justify the organization of the solitons in branches. Section~\ref{Sec:Numerical:Arguments}
introduces a useful algorithm that can be used to locate all the
branches of solitons and shows that $p=0$ solitons (and the
continuous spectrum) only exist in the small amplitude limit.
These results are extended to plasmas with mobile ions in
the Appendix. In Sec.~\ref{Sec:Quasisoliton} a numerical algorithm
to locate exact solutions is formulated and used to locate
new families of quasi-solitons. It is shown that the
branches of $p\neq 0$ solitons are parametrically embedded inside
the continuous spectrum of quasi-solitons.
The stability of these structure is explored in Sec.~\ref{Sec:Quasi:Stability}
and our conclusion are presented in Sec.~\ref{Sec:Conclusions}.

\section{The fluid model\label{Sec:Model}}

We consider a plasma consisting of electrons and immobile ions. For
convenience, length, time, velocity, momentum, vector and scalar
potentials and density are normalized by $c/\omega_{pe}$,
$\omega_{pe}^{-1}$, $c$, $m_ec$, $m_ec^2/e$ and $n_0$, respectively.
Here $n_0$, $\omega_{pe}=\sqrt{4\pi n_0e^2/m_e}$, $m_e$ and $c$ are
the unperturbed plasma density, the electron plasma frequency, the
electron mass and the speed of light. Maxwell (in the Coulomb gauge)
and plasma equations then read in the laboratory frame $S$
\begin{subequations}
\begin{equation}
\Delta \bm{A}-\frac{\partial^2\bm{A}}{\partial
t^2}-\frac{\partial}{\partial t}\nabla\phi =
\frac{n}{\gamma}\bm{p}\,,\label{Eq:A}
\end{equation}
\begin{equation}
\Delta \phi = n-1\,,\label{Eq:Poisson}
\end{equation}
\begin{equation}
\frac{\partial n}{\partial
t}+\nabla\cdot\left(n\bm{v}\right)=0\,,\label{Eq:Continuity}
\end{equation}
\begin{equation}
\frac{\partial \bm{P}}{\partial
t}-\bm{v}\times\left(\nabla\times\bm{P}\right)=\nabla\left(\phi-\gamma\right)\,,\label{Eq:P}
\end{equation}\label{Sys:Fluid:0}
\end{subequations}
where $\bm{A}$ and $\phi$ are the vector and scalar potentials, $n$ is the electron plasma density,
$\bm{P}=\bm{p}-\bm{A}$, $\gamma=\sqrt{1+|\bm{p}|^2}$ and $\bm{p}$
and $\bm{v}=\bm{p}/\gamma$ are  the electron momentum and velocity,
respectively.

Assuming 1-dimensional ($\partial/\partial y=\partial/\partial z = 0$)
solitons, one finds from \refeq{Sys:Fluid:0} that $A_x =
0$, $P_y = P_z = 0$ and
\begin{subequations}
\begin{equation}
\frac{\partial^2 A_{y,z}}{\partial x^2}-\frac{\partial^2
A_{y,z}}{\partial t^2} = \frac{n}{\gamma}A_{y,z}\,,\label{Eq:AyAz}
\end{equation}
\begin{equation}
\frac{\partial^2 \phi }{\partial t\partial x}=-\frac{n}{\gamma}p_x\,,
\label{Eq:px}
\end{equation}
\begin{equation}
 n = 1+\frac{\partial^2\phi}{\partial x^2}\,,\label{Eq:n}
\end{equation}
\begin{equation}
 \frac{\partial \phi}{\partial x} = \frac{\partial \gamma}{\partial x}+\frac{\partial p_x}{\partial t}\,,\label{Eq:ex}
\end{equation}
\begin{equation}
\gamma = \sqrt{1+p_x^2+A_y^2+A_z^2}\,.\label{Eq:gamma}
\end{equation}\label{Sys:Fluid:1d}
\end{subequations}

We are interested in solutions of the form
\begin{equation}\label{eq:Ansatz}
A_y+iA_z = a(\xi)e^{-i\omega\tau}\,,
 \end{equation}
 with
 \begin{equation}
 \xi\equiv
\frac{x-Vt}{\sqrt{1-V^2}}\,, \ \ \ \ \tau\equiv
\frac{t-Vx}{\sqrt{1-V^2}}\,,
\end{equation}
$\phi=\phi(\xi)$, $n=n(\xi)$, and $\gamma=\gamma(\xi)$ and boundary
conditions $a\rightarrow 0$, $\phi\rightarrow 0$, $n\rightarrow 1$
and $\gamma\rightarrow 1$ as $\xi\rightarrow \infty$ (or
$\xi\rightarrow -\infty$ ). Under these assumptions \refeq{Eq:AyAz} and \refneq{Eq:gamma}
become two ordinary differential
equations \cite{Kozlov_79,Kaw_92,Farina01a}
\begin{subequations}
\begin{equation}
a''=\left(\frac{V}{R_e}-\omega^2\right)a\,\label{Eq:Circular:a}
\end{equation}
\begin{equation}
\phi''=V\left(\frac{\psi_e}{R_e}-\frac{1}{V}\right)\,\label{Eq:Circular:phi}
\end{equation}
\end{subequations}
where the prime denotes derivative with respect to $\xi$ and we
introduced the auxiliary functions
\begin{equation}
R_e(a,\phi;V)\equiv \sqrt{\psi_e^2-(1-V^2)(1+a^2)}\,\label{Eq:R}
\end{equation}
and
\begin{equation}
\psi_e \equiv 1+\phi\,.
\end{equation}
Once $a$ and $\phi$ are known, the fluid variables are computed from
\begin{align}
  p_x &= (V\psi_e-R_e)/(1-V^2),\label{eq:fluid:vars}\\
  \gamma_e &=(\psi_e-VR_e)/(1-V^2),\\
  n &= V(\psi_e/R_e-V)/(1-V^2).\label{eq:fluid:vars:Final}
\end{align}

Equations \refeqs{Eq:Circular:a}{Eq:Circular:phi} have several
properties that are useful when discussing the existence of solitary
waves. First, they result from the Hamiltonian
\begin{align}
H(a,p_a,\phi,p_\phi)=&
\frac{1-V^2}{2}\left[\left(\frac{p_a}{1-V^2}\right)^2+\omega^2a^2\right]\nonumber\\
&-\frac{1}{2}p_\phi^2+V R_e(a,\phi)-\psi_e
\end{align}
where the momenta are $p_a=(1-V^2)a'$ and $p_\phi=-\phi'$. Since $H$
does not depend on $\xi$ explicitly, it is a conserved quantity. For our
boundary conditions we have $H=V^2-1$.
In addition, the system is reversible in the sense defined by Devaney
\cite{Devaney76}: there is a reversing involution $G_j$ (a discrete symmetry operation)
which fixes half the phase variables and under which the system is invariant
under $\xi$-reversal ($\xi\rightarrow -\xi$). In other words,
writing Eqs.~(\ref{Eq:Circular:a})-(\ref{Eq:Circular:phi}) as $\bm{x}_s'
= \bm{f}(\bm{x}_s)$ with $\bm{x}_s \equiv [a\ a'\ \phi\ \phi']\in
\mathbb{R}^{4}$, there is an involution $G_j$ that satisfies
\begin{equation}
G_j\bm{f}(\bm{x}_s)=-\bm{f}(G_j\bm{x}_s), \ \ \ G_j^2=\mathds{1}, \ \
\ \dim(S_j)=2\,,
\end{equation}
where the subspace $S_j=Fix(G_j):=\left\{\bm{x}_s:
G_j\bm{x}_s=\bm{x}_s\right\}$ is called the symmetric section of
reversibility. Equations \ref{Eq:Circular:a}-\ref{Eq:Circular:phi}
have two involutions. The first one is
\begin{equation}
G_1:\ (a,a',\phi,\phi')\rightarrow
(a,-a',\phi,-\phi')\,,\label{Eq:Involution:1}
\end{equation}
with symmetric section
\begin{equation}\label{Eq:S1}
 S_1:\quad a'=\phi'=0\,,
\end{equation}
and the second
\begin{equation}
G_2:\ (a,a',\phi,\phi')\rightarrow
(-a,a',\phi,-\phi')\,,\label{Eq:Involution:2}
\end{equation}
with symmetric section
\begin{equation}\label{Eq:S2}
  S_2:\quad a=\phi'=0\,.
\end{equation}
Note also that the subspace $a=a'=0$, corresponding to pure
electrostatic excitations, is invariant.

\refEqs{Eq:Circular:a}{Eq:Circular:phi} are singular
when $V=0$. For such a case, one readily finds from \refeq{Eq:px}~and~\refneq{Eq:gamma}
that $p_x=0$ and $\gamma=\sqrt{1+a^2}$. Using
these results in \refeq{Eq:ex} gives the invariant
\begin{equation}
R_e(a,\phi;0) = 0\label{Eq:R:V0}\,,
\end{equation}
that provides a relation between $\phi$ and $a$. A differential equation for the latter
is found by combining \refeq{Eq:AyAz} and \refneq{Eq:n} to yield
\begin{equation}
a''=a\left(1+a^2\right)\left(\frac{1}{\sqrt{1+a^2}}-\omega^2\right)+\frac{aa'^2}{1+a^2}\,.\label{Eq:a:V0}
\end{equation}
\refEq{Eq:a:V0} has the first integral
\begin{equation}
H
=\frac{1}{2}\left(\frac{a'^2}{1+a^2}+\omega^2a^2\right)-\sqrt{1+a^2}\label{Eq:H:V0}\,,
\end{equation}
and also admits the two involutions $G_1:\ (a,a')\rightarrow
(a,-a')$ and $G_2:\ (a,a')\rightarrow (-a,a')$. \refEq{Eq:a:V0}
admits the solitary wave solution \cite{Esirkepov_02}
\begin{equation}
a(\xi)=\frac{2\sqrt{1-\omega^2}\cosh\left(\xi\sqrt{1-\omega^2}\right)}{\cosh^2\left(\xi\sqrt{1-\omega^2}\right)-\left(1-\omega^2\right)}\,,\label{Eq:Esirkepov}
\end{equation}
with maximum amplitude
\begin{equation}
A_{\mathrm{max}}=\frac{2\sqrt{1-\omega^2}}{\omega^2}\,.
\end{equation}
These standing solutions exist for $\sqrt{2/3}<\omega<1$.
The soliton with maximum amplitude, exhibiting
zero electron density at its center, has $\omega=\sqrt{2/3}$.

\section{Organization of solitons in parameter space\label{Sec:Solitary:Waves}}

\subsection{Geometric Arguments \label{Sec:Geometric:Arguments}}
Solitons of the cold fluid model described by Sys.
(\ref{Sys:Fluid:1d}) can be found as homoclinic orbits of
\refeqs{Eq:Circular:a}{Eq:Circular:phi}, i.e. solutions that are
asymptotic to a fixed point $Q_0$ in both $\xi\rightarrow \pm
\infty$. Even without computing them analytically or numerically,
their organization in parameter space can be anticipated by using
arguments based on the dimension of the phase space, the number of
conserved quantities, reversibility properties, and the dimensions
of the stable and unstable manifolds of the fixed point. We recall
that the stable $W^s$ (unstable $W^u$) manifold of a fixed point is
the set of forward (backward) in $\xi$ trajectories that terminate
at the fixed point. Since the solitons are homoclinic orbits that
connect to $Q_0$ as $\xi\rightarrow \pm \infty$, these orbits lie in
the intersection of $W^s$ and $W^u$. Moreover, the solitons studied
here are invariant under the $\xi$-reversion transformations $G_1$
or $G_2$ [\refeqs{Eq:Involution:1}{Eq:Involution:2}] and this
implies that $W^s$ and $W^u$ have to intersect on one of the
sections $S_1$ or $S_2$ defined by \refeqs{Eq:S1}{Eq:S2}.
In \refref{Farina01a} the solitons are clasified
according to the number of zeros $p$ of the
vector potential (number of nodes). Solitons with an even (odd) number of nodes
are invariant under $G_1$ ($G_2$) and thus
intersect with $S_1$ ($S_2$). This later result will be
used hereafter to discuss the existence of solitons. Interested
readers can find an excellent review of the theory in \refref{champneys1998}.

Let us first examine the case $V\neq 0$. Linearizing \refeqs{Eq:Circular:a}{Eq:Circular:phi}
about the fixed point $Q_0=(a=0,a'=0,\phi=0,\phi'=0)$ gives
\begin{equation}
  a''-(1-\omega^2)a = 0\label{Eq:a:linearized}
\end{equation}
\begin{equation}
  \phi''+\frac{1-V^2}{V^2}\ \phi = 0\label{Eq:phi:linearized}
\end{equation}
Looking for solutions of the form $\exp(\lambda\xi)$, one finds that
Eqs. \ref{Eq:a:linearized} and \ref{Eq:phi:linearized}  have
eigenvalues $\lambda_{1,2}=\pm\sqrt{1-\omega^2}$ and
$\lambda_{3,4}=\pm\sqrt{-(1-V^2)/V^2}$, respectively. Hereafter we
restrict the analysis to $\omega<1$, when the fixed point is a
saddle-center and solitons can exist. For such a case, the stable
$W^s$ and unstable $W^u$ manifolds of the fixed point are
one-dimensional.
By $\xi$-reversibility one can see that if the unstable manifold $W^u$
intersects any of the symmetric sections $S_1$ or $S_2$, then
the stable manifold $W^s$ has to also intersect it at the same point
(in fact, this implies that $W^s=W^u$ since both manifolds are one-dimensional).
Therefore, the condition for existence of a homoclinic orbit is that
the one-dimensional $W^u$ (or $W^s$) intersect the two-dimensional section $S_i$
within the three-dimensional energy shell $H=V^2-1$.
In general~\cite{champneys1998}, such an intersection is expected to occur for specific parameter values
that form branches in the $V$-$\omega$ parameter space.

For $V=0$, the dynamics is governed by \refeq{Eq:a:V0} and the
phase space is two-dimensional. One readily verifies that
$Q_0=(a=0,a'=0)$ is a fixed point with eigenvalues
$\lambda_{1,2}=\pm\sqrt{1-\omega^2}$. If $\omega<1$, then $Q_0$ is a
saddle and the dimensions of $W^s$ and $W^u$ are both equal to 1.
The intersection of the one-dimensional unstable manifold $W^u$ with
the one-dimensional  section $a'=0$ within the two-dimensional
$(a,a')$ phase space is robust against parameter variations. This implies
that $V=0$ solitons exist within a continuous range of $\omega$.
This is in agreement with the analytical solution
[\refeq{Eq:Esirkepov}] that exists within the frequency domain
$2/3<\omega^2<1$.

The previous discussion highlights the crucial difference between
standing and moving solitons: the existence of the invariant
$R(a,\phi,0)$ if $V=0$ [\refeq{Eq:R:V0}] which restricts the
dimensionality of phase-space and leads to a continuous spectrum in
that case. Equations (\ref{Eq:Circular:a})-(\ref{Eq:Circular:phi})
share a set of properties that are compatible with the hypothesis
assumed in Ref.~\cite{Mielke_92}:  Hamiltonian structure,
reversibility, saddle-center fixed point and the invariant subspace
$a=a'=0$. In addition to the cascade of homoclinic orbits and a
rigorous investigation of the structure of the set made by parameter
values yielding localized structures, Ref.~\cite{Mielke_92}
also proved the existence of families of periodic orbits and Smale
horseshoes in nearby level sets of the Hamiltonian function. The
latter proves that the system is not integrable globally. For this
reason  a global invariant in the case $V\neq 0 $ that could play
the role of \refeq{Eq:R:V0} for $V = 0$ is not possible.

\subsection{Numerical method to locate solitons \label{Sec:Numerical:Arguments}}

This section introduces a procedure to locate solitons and explains
why past works concluded that there is a region in the
$V-\omega$ plane with a continuous spectrum. For any given value of the
parameters $V$ and $\omega$ we carried out integrations of Eqs.
\ref{Eq:Circular:a} and \ref{Eq:Circular:phi} with a symplectic
fourth order Runge-Kutta-Nystrom method. We took an initial
condition that belongs to the linear approximation of the unstable
manifold of $Q_0$. This approximation reads
\begin{equation}
  \bm{x}_s(\xi=0)=\frac{\epsilon}{\sqrt{2-\omega^2}}\left[1 \ \
  \sqrt{1-\omega^2}\ \ 0 \ \ 0 \right]\label{Eq:IC}
\end{equation}
with $\epsilon$ a small parameter. The integrations were stopped at
$\xi=\xi_s$, when the orbit intersects the section $\phi'=0$ with a
tolerance smaller than $10^{-15}$. The algorithm used a fixed
$\xi$-step ($\Delta\xi$), except at the last point of the orbit. For
the latter, the Newton-Raphson method was used to find a $\Delta\xi$
such that $\phi'(\xi_s)=0$ was satisfied with the prescribed
tolerance. All the calculations were carried with $\Delta\xi=5\times
10^{-4}$ and $\Delta\xi= 5\times 10^{-5}$ in order to check the
convergence of the results.

\begin{figure}[h]
\noindent\includegraphics[width=20pc]{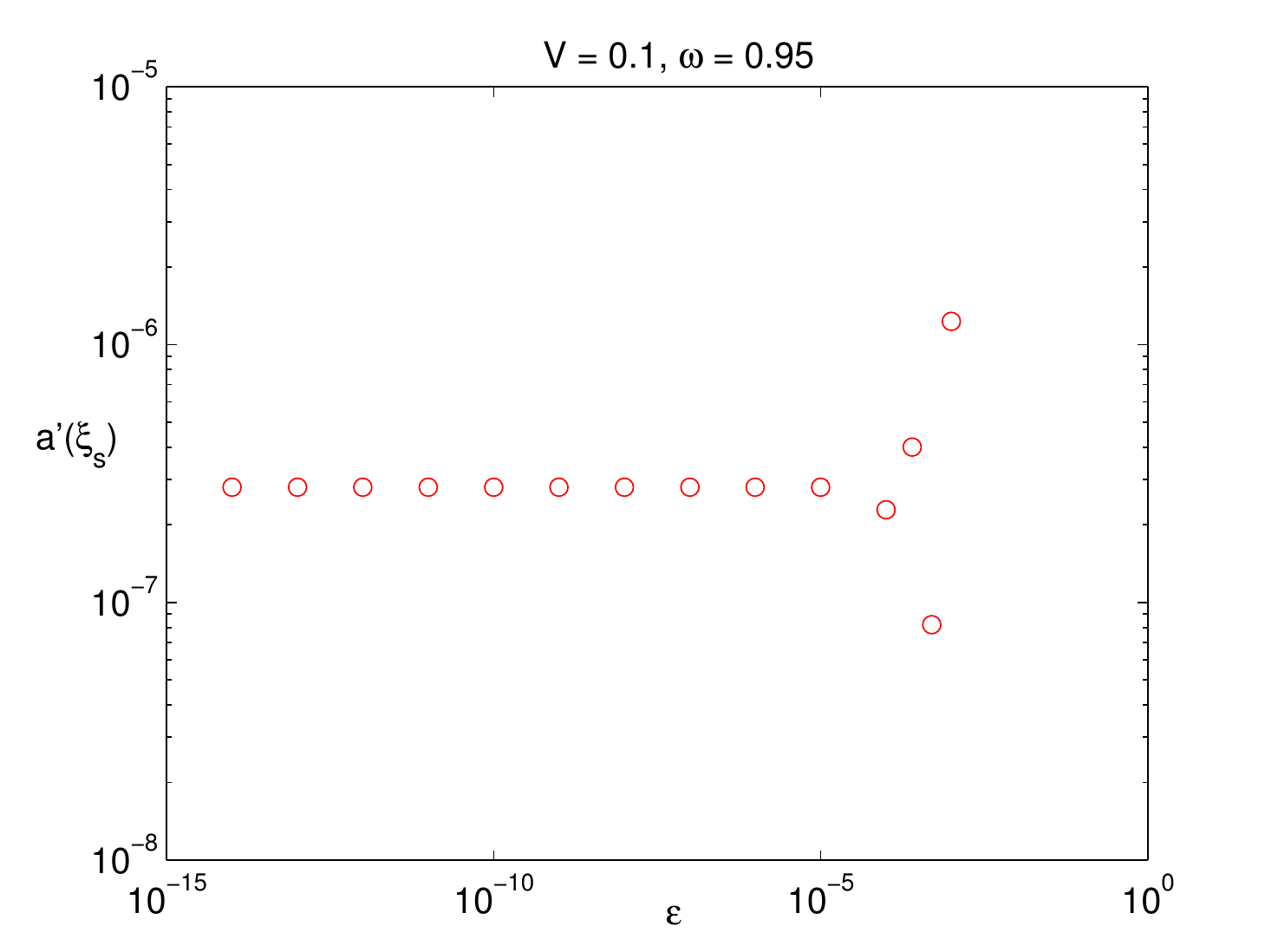}
\caption{Value of $a'(\xi_s)$ versus $\epsilon$ for integrations
started with Eq. \ref{Eq:IC} and $\xi_s$ given by
$\phi'(\xi_s)=0$\label{Fig:a_versus_T}}
\end{figure}

Figure \ref{Fig:a_versus_T} shows the magnitude of $a'(\xi_s)$
versus $\epsilon$ for integrations started with the initial
condition \refeq{Eq:IC} and parameter values ($V=0.1$ and
$\omega=0.95$) that belong to the domain where a continuous spectrum
has been predicted. As shown in Fig.~\ref{Fig:a_versus_T}, for
$\epsilon\lesssim 10^{-5}$ the distance from the symmetric section
$S_1$ remains constant. If a soliton existed for this set of
parameters, the magnitude of $a'(\xi_s)$ should vanish as
$\epsilon\rightarrow 0$. We remark that any computational method
with a tolerance above $10^{-6}$ would interpret this orbit, which
is not homoclinic, as a soliton.

\begin{figure}[h]
\noindent\includegraphics[width=20pc]{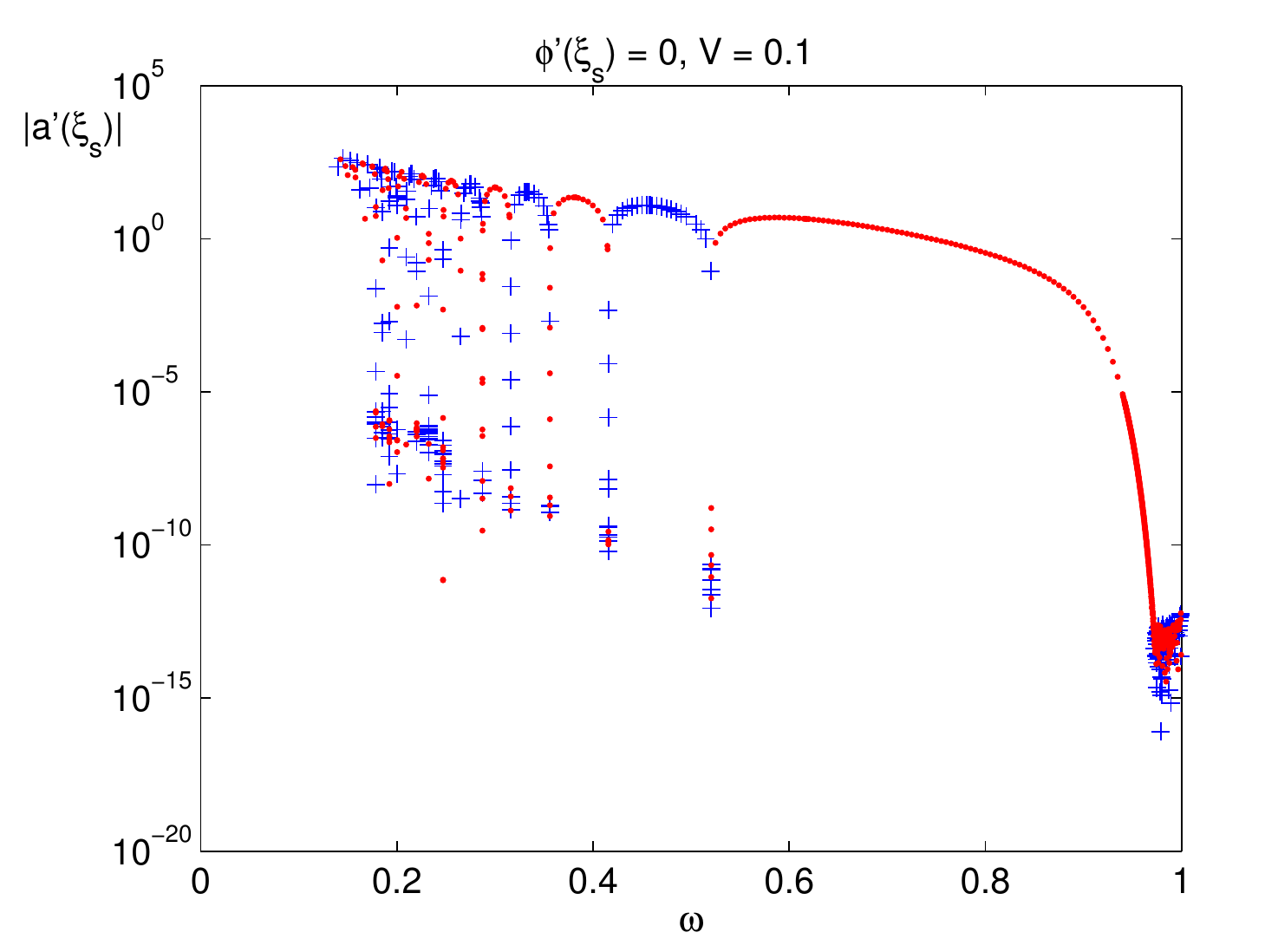}
\caption{Absolute value of $a'(\xi_s)$ versus $\omega$ for $V =
0.1$. Crosses (dots) corresponds to positive (negative) values of
$a'(\xi_s)$.\label{Fig:ap_versus_omega}}
\end{figure}

In order to investigate further the organization of the solitary
waves in the $V-\omega$ plane we fixed $V = 0.1$, and carried out
integrations for several $\omega$ values and initial conditions
given by \refeq{Eq:IC} with $\epsilon=10^{-7}$. Again we stopped
the integration at $\xi=\xi_s$, i.e. when the orbit intersected the
section $\phi'=0$. At each $\omega$ value we recorded the value of
$a'(\xi_s)$.
Figure \ref{Fig:ap_versus_omega} shows
the absolute value of $a'(\xi_s)$ in logarithm scale versus
$\omega$. Crosses (dots) were used to present positive (negative)
values of $a'(\xi_s)$. The figure shows that, for $\omega<0.6$, the sign of $a'(\xi_s)$
changes at several $\omega$ values, with the $\omega$ interval
between two sign changes becoming smaller as $\omega$ decreases.
Obviously, at each change of sign, there is an $\omega$ value, say $\omega_j$,
that makes $a'(\xi_s)=0$ and this designates a trajectory that intersects the
section $S_1$ (p is even), i.e. a soliton. Since solitons occur at discrete values
of $\omega$ for a given $V$, the solitons are organized in branches
in the $\omega-V$ plane.

\begin{figure}[h]
\noindent\includegraphics[width=20pc]{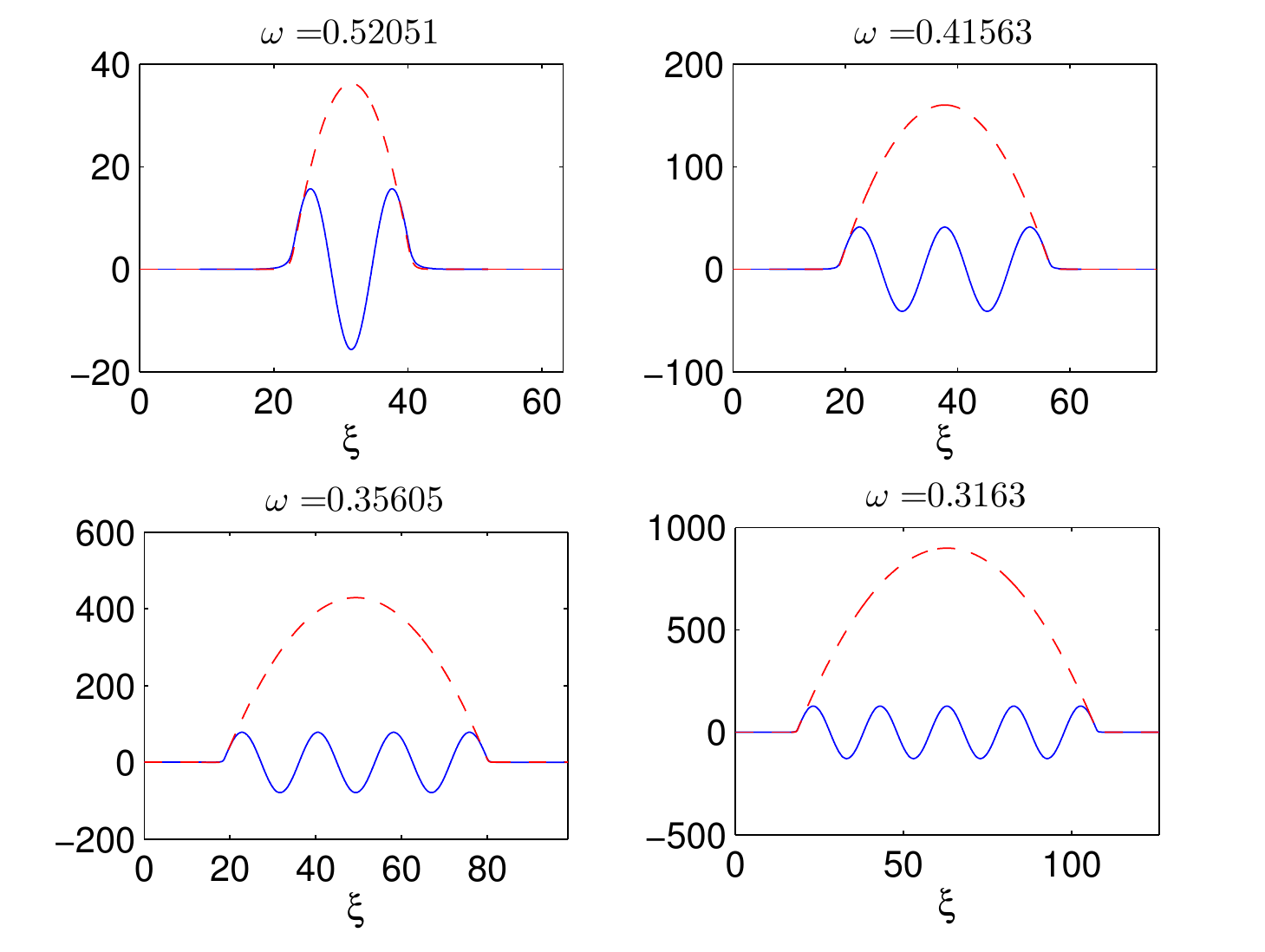} \caption{Some
examples of solitons that intersect the section $S_1$ with $V=0.1$.
Solid and dashed lines correspond to $a$ and $\phi$, respectively.
\label{Fig:Soliton:Par}}
\end{figure}

The diagram shown in Fig.  \ref{Fig:ap_versus_omega}  can be used to
compute the branches with even $p$. Once a change of sign of
$a'(\xi_s)$ is detected, the value of  $\omega_j$ can be computed
easily with a bisection method. Figure \ref{Fig:Soliton:Par}
displays four examples at $V = 0.1$ computed with this method. Each
soliton in Fig. \ref{Fig:Soliton:Par} belongs to a different branch.
Branches with lower value of $\omega_j$ have a vector potential with
a higher number of zeros  and also much higher amplitudes.

\begin{figure}[h]
\noindent\includegraphics[width=20pc]{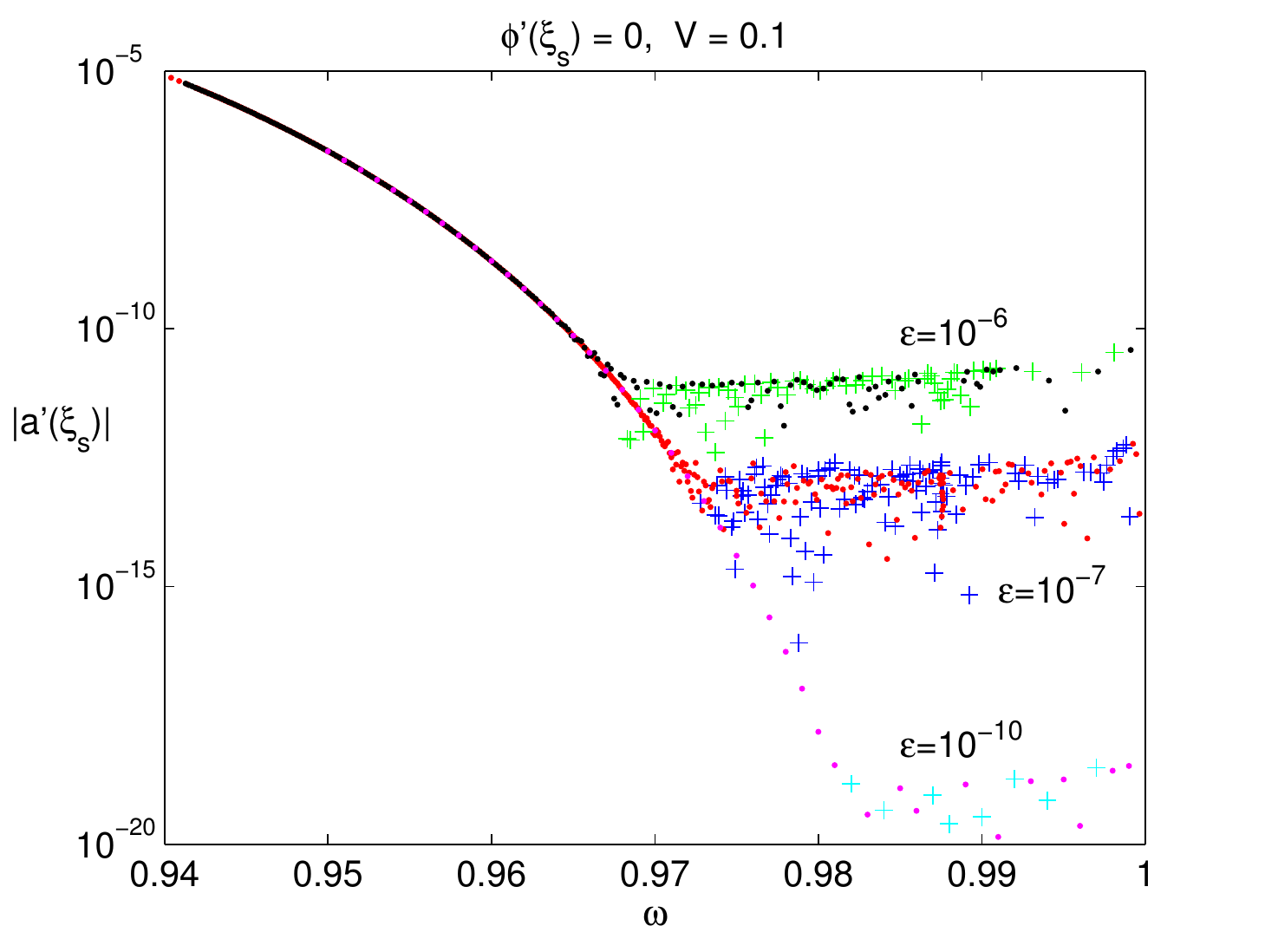}
\caption{Absolute value of $a'(\xi_s)$ versus $\omega$ for $V = 0.1$
and three different values of $\epsilon$ in Eq.~(\ref{Eq:IC}). Crosses
(dots) corresponds to positive (negative) values of $a'(\xi_s)$.
\label{Fig:ap_versus_omega_Zoom}}
\end{figure}

As $\omega$ approaches to $1$, $|a'(\xi_s)|$  decreases
monotonically (see Fig. \ref{Fig:ap_versus_omega}) but we do not
observe a clear indication of an intersection with the symmetric
section $S_1$. Beyond a certain $\omega$, say $\omega^*$, the
monotonic decrease of $|a'(\xi_s)|$ appears to be disrupted and we
observe erratic changes in sign of $a'(\xi_s)$. A detail of this
part of the diagram is shown in Fig.~\ref{Fig:ap_versus_omega_Zoom},
which contains calculations with $\epsilon=10^{-6}$ and
$\epsilon=10^{-7}$. We find that the onset of this effect appears at
a larger $\omega^*$ for smaller $\epsilon$ and therefore is a
consequence of the truncation of the infinite domain
$-\infty<\xi<\infty$ in our calculation, i.e. the finite value of
$\epsilon$ in Eq. \ref{Eq:IC}.
The numerical integrations were started at a distance $\epsilon$
from $Q_0$ and along a linear approximation of the unstable
manifold. For this reason, we cannot expect that the behavior of
$|a'(\xi_s)|$ versus $\omega$ will be smooth for $|a'(\xi_s)|\sim
\epsilon^2$.
For instance, for $\epsilon=10^{-6}$ and $\epsilon=10^{-7}$ the
erratic behavior starts when $|a'(\xi_s)|$ reaches about $10^{-12}$
and $10^{-14}$, respectively.
In order to investigate the dynamics with $\omega$ even closer to 1,
we need to reduce $\epsilon$ more but then we are limited by the
finite precision in our computations. To overcome this obstacle we
use the arbitrary precision capabilities of Mathematica in order to
perform computations with 30 digits of precision. We use the
built-in adaptive symplectic integrator with an (absolute) error
tolerance of 25 digits. This allows to verify the results obtained
with $\epsilon=10^{-6},\, 10^{-7}$ with an independent code and also
to perform computations with $\epsilon=10^{-10}$ as shown in
\reffig{Fig:ap_versus_omega_Zoom}. We observe that the onset of
erratic behavior of the sign of $a'(\xi_s)$ is now at even larger
$\omega^*$ while $|a'(\xi_s)|$ becomes of the order of $10^{-19}$.

The trend of the $a'(\xi_s)-\omega$ curve and the geometric
arguments of Sec.~\ref{Sec:Geometric:Arguments} seem to rule out the
existence of a continuous spectrum of finite amplitude $p=0$
solitons. In particular, there appears to be no
$\epsilon$-independent $\omega$ value signifying transition to such
a continuous spectrum. What the behavior of $a'(\xi_s)-\omega$ as
$\omega\rightarrow1$ seems to indicate is that $p=0$ solitons exist
in the small amplitude limit. Indeed, in the limit of
$\omega\rightarrow1$ (small solution amplitude) a nonlinear
Schr\"odinger equation (NLS) limit of the Maxwell-fluid model
exists~\cite{siminos2014} 
which supports soliton solutions. However, one has to note that the NLS
equation is integrable, possesses infinitely many integrals of
motion, thus naturally leading to the existence of solitons.

\begin{figure}[h]
\noindent\includegraphics[width=20pc]{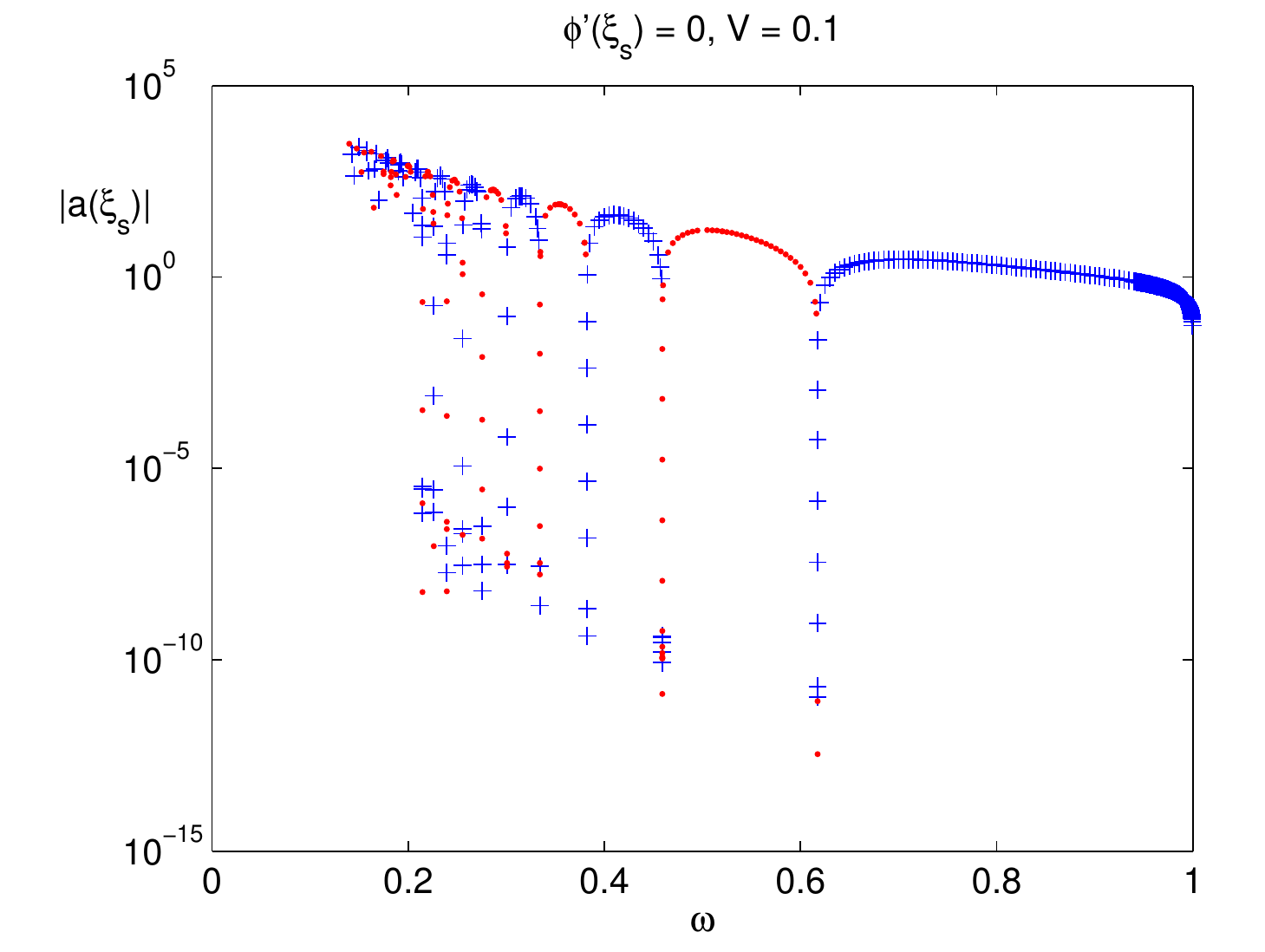}
\caption{Absolute value of $a(\xi_s)$ versus $\omega$ for $V = 0.1$.
Crosses (dots) corresponds to positive (negative) values of
$a(\xi_s)$.\label{Fig:a_versus_omega}}
\end{figure}

\begin{figure}[h]
\noindent\includegraphics[width=20pc]{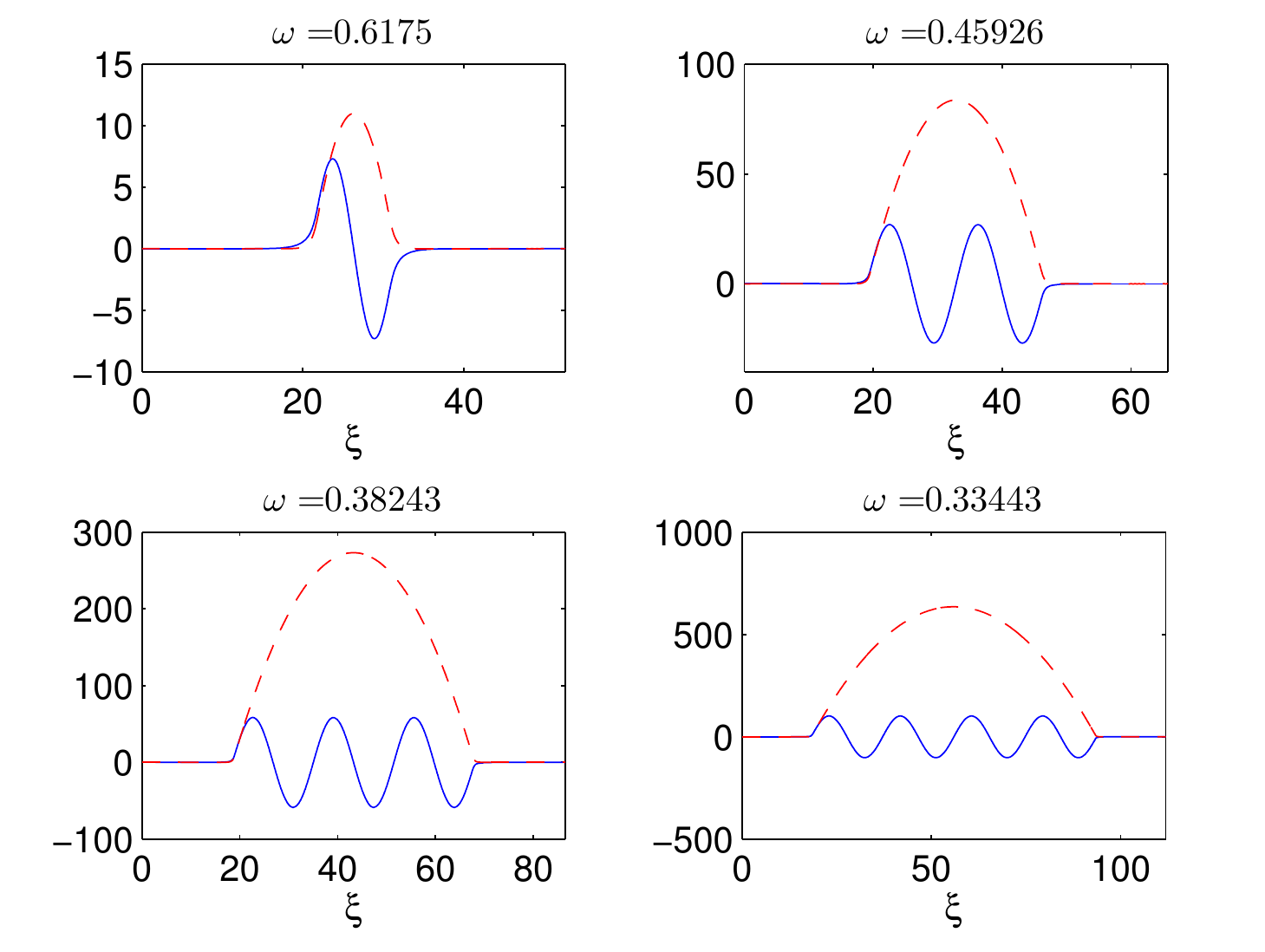} \caption{Some
examples of solitons that intersect the section $S_2$ with $V=0.1$.
Solid and dashed lines correspond to $a$ and $\phi$,
respectively.\label{Fig:Soliton:Impar}}
\end{figure}

Plotting the absolute value of $a(\xi_s)$ (instead of $a'(\xi_s)$)
versus $\omega$ in logarithm scale and using crosses and dots to
denote positive and negative values of $a(\xi_s)$ reveals the
organization of the solitons with odd $p$ (see Fig.
\ref{Fig:a_versus_omega}). These are orbits that intersect the $S_2$
section. There are many changes of sign of $a(\xi_s)$ and, at the
particular $\omega_j$ values that make $a(\xi_s)=0$, there are
branches of solitons with an odd number of nodes in the vector potential.
Some of these orbits are shown in Fig. \ref{Fig:Soliton:Impar}.

Tables \ref{Table:Par} and \ref{Table:Impar} summarize some of the
numerical results for $V = 0.1$. It contains branches of solutions
with the number of zeros of the vector potential $p$ from 1 to 20 as
well as the frequency values $\omega_j$ which give rise to localized
solitons.

 \begin{table}[h]
 \caption{Solitons spectrum with $V = 0.1$ (even $p$).\label{Table:Par}}
 \begin{ruledtabular}
 \begin{tabular}{c c c  }
 p &  $\phi_{max}$ & $\omega_j$\\
\hline
2 &  $3.629779\times 10^{1}$  & 0.52050770\\
4 &  $1.603032\times 10^{2}$  & 0.41562715\\
6 &  $4.290256\times 10^{2}$  & 0.35604770\\
8 &  $8.990913\times 10^{2}$  & 0.31629641\\
10 & $1.627378\times 10^{3}$ & 0.28736027\\
12 & $2.670918\times 10^{3}$ & 0.26509070\\
14 & $4.086845\times 10^{3}$ & 0.24727222\\
16 & $5.932369\times 10^{3}$ & 0.23259912\\
18 & $8.264754\times 10^{3}$ & 0.22024540\\
20 & $1.114131\times 10^{4}$ & 0.20965960\\

 \end{tabular}
 \end{ruledtabular}
 \end{table}

 \begin{table}[h]
 \caption{Solitons spectrum with $V = 0.1$ (odd $p$).\label{Table:Impar}}
 \begin{ruledtabular}
 \begin{tabular}{c c c  }
 p &  $\phi_{max}$ & $\omega_j$\\
\hline
1  & $1.099531\times 10^{1}$  & 0.61749714\\
3  & $8.374166\times 10^{1}$  & 0.45926068\\
5  & $2.730471\times 10^{2}$  & 0.38242601\\
7  & $6.353430\times 10^{2}$  & 0.33443103\\
9  & $1.227398\times 10^{3}$  & 0.30080027\\
11 & $2.106175\times 10^{3}$  & 0.27555955\\
13 & $3.328761\times 10^{3}$  & 0.25572137\\
15 & $4.952331\times 10^{3}$  & 0.23960230\\
17 & $7.034125\times 10^{3}$  & 0.22617167\\
29 & $9.631429\times 10^{3}$  & 0.21475857\\

 \end{tabular}
 \end{ruledtabular}
 \end{table}

\section{Quasi-solitons and embedded solitons\label{Sec:Quasisoliton} }

As we now show, the solitons computed in
Sec.~\ref{Sec:Solitary:Waves} are special members of a new family of
delocalized solutions named quasi-solitons. The latter do not only
provide a broader understanding of the former but also have an
interesting physical meaning. In general, quasi-soliton are
weakly non-localized solutions with a soliton-like core and non-vanishing
oscillatory tails~\cite{boyd_1998}. In the present case, the
quasi-solitons that we will find are partially localized: the
electromagnetic field is localized but the longitudinal quantities,
e.g. $n$ and $\phi$ exhibit oscillatory tails.

From a dynamical system point of view, a quasi-soliton is an
homoclinic orbit that connects at $\xi\rightarrow \pm\infty $ with a
periodic orbit. In a reversible Hamiltonian system like
Eq.~(\ref{Eq:Circular:a})-(\ref{Eq:Circular:phi}), the existence of
a quasi-soliton requires the intersection of the 2-dimensional
unstable manifold of the periodic orbit with the 2-dimensional
symmetric section. Hence, symmetric quasi-solitons are persistent
under parameter variations and they would appear in a continuous
region of the $\omega-V$ plane~\cite{champneys1998}. As shown in the
context of nonlinear optics \cite{Yang_99}, the amplitudes of the
oscillating tails of the quasi-solitons can exactly vanish at a
discrete set of parameter values. In other words, for certain
relations $V=V(\omega)$ these solutions are truly localized and
correspond to branches of true solitons. For this reason, such
solitons are said to be parametrically \emph{embedded} in the family
of delocalized quasi-soliton.

Before we explore if this mechanism also applies in our problem
we note that since we found Eqs.~(\ref{Eq:Circular:a})-(\ref{Eq:Circular:phi})
from Eqs.~(\ref{Eq:AyAz})-(\ref{Eq:gamma}) by imposing the boundary
conditions $a\rightarrow 0$, $\phi\rightarrow 0$, $n\rightarrow 1$
and $\gamma\rightarrow 1$ as $\xi\rightarrow \infty$, only solutions
of  Eqs.~(\ref{Eq:Circular:a})-(\ref{Eq:Circular:phi}) connecting with
the equilibrium point $Q_0$ correspond to solutions of the fluid
system given by Eqs.~(\ref{Eq:AyAz})-(\ref{Eq:gamma}). However,
quasi-solitons connect with a periodic orbit and not with $Q_0$.
For this reason, we will compute quasi-soliton solutions directly from
Eqs.~(\ref{Eq:AyAz})-(\ref{Eq:gamma}) with an extension to $V\neq 0$ of
the finite-difference algorithm presented in Ref.~\cite{Sanchez_2015}.

\subsection{Numerical algorithm to locate quasi-solitons\label{Sec:Quasi:Algorithm}}

Since this work deals with traveling waves, we introduce a
boosted frame $S'$ that moves with constant velocity $V$ along the
$x$-axis. Four-vectors $(t,\bm{x})$, $(\gamma,\bm{p})$,
$(\phi,\bm{A})$ and $(\rho,\bm{j})$ with the electron charge and
current densities given by $\rho=-n$ and $\bm{j}=-n\bm{v}$ transform
according to Lorentz transformations. Equations
\refeqs{Eq:AyAz}{Eq:gamma} in the boosted frame read
\begin{subequations}
\begin{equation}
\frac{1}{L^2}\frac{\partial^2 A'_{y,z}}{\partial
x'^2}-\Omega^2\frac{\partial^2 A'_{y,z}}{\partial t'^2} =
\frac{n'}{\gamma'}A'_{y,z}\label{Eq:Moving:AyAz}
\end{equation}
\begin{equation}
\Omega\frac{\partial e_x' }{\partial t'}-\frac{V}{L}\frac{\partial
e_x' }{\partial x'}=\frac{n'}{\gamma'}\left(p_x'+V\gamma'\right)
\label{Eq:Moving:px}
\end{equation}
\begin{equation}
 n' =\frac{\gamma'}{\gamma'+Vp_x'}\left[\sqrt{1-V^2}-\left(\frac{1}{L}\frac{\partial e_x'}{\partial x'}-\Omega V\frac{\partial e_x'}{\partial t'}\right)\right]\label{Eq:Moving:n}
\end{equation}
\begin{equation}
 e_x' = -\left(\frac{1}{L}\frac{\partial \gamma'}{\partial x'}+\Omega \frac{\partial p_x'}{\partial t'}\right)\label{Eq:Moving:ex}
\end{equation}
\begin{equation}
\gamma' = \sqrt{1+p_x'^2+A_y'^2+A_z'^2}\label{Eq:Moving:gamma}
\end{equation}\label{sys:fluid:boosted}
\end{subequations}
where we used a prime to denote the variables in the boosted frame
and introduced the normalized distance and time
\begin{equation}
x' = \frac{1}{L}\frac{x-Vt}{\sqrt{1-V^2}}\ \ \ \ \ \ \ \ \ \ t'
=\Omega \frac{t-Vx}{\sqrt{1-V^2}},
\end{equation}
For numerical convenience we also introduced the parameters $L$ and
$\Omega$, which allow to work in a computational box $x'\in [0,\ 1]$
and $t'\in [0, \ 2\pi]$.

Quasi-solitons are spatially- and temporally-periodic solutions of
\refeq{sys:fluid:boosted} with unknown spatial period $L$ and
angular frequency $\Omega$. We note that we do not impose a
restriction on the functional form of these solutions as we did in
writing \refeq{eq:Ansatz}. In our algorithm the computational box is
discretized  with $N_x$ and $N_t$ regularly spaced points. A vector
of unknowns $\bm{b}=[A_{yi,j} \ \ A_{zi,j}\ \ p_{xi,j}\ \ n_{i,j}\ \
e_{xi,j}\ \ \gamma_{i,j} ]$ with $i=1,...N_x$ and $j = 1,...N_t$
with the values of the variables at the grid points is constructed.
After using second order finite-difference formulae to approximate
the spatial and temporal derivatives and imposing periodic boundary
conditions in space and time, one finds from
\refeq{sys:fluid:boosted} a large set of nonlinear algebraic
equations of the form
\begin{equation}\label{eq:algebr:quasisol}
\bm{g}(\bm{b};V,\Omega,L)=0\,
\end{equation}
for the unknowns $\bm{b}$ ($V,\,\Omega$ and $L$ are treated as fixed
parameters). This is solved iteratively with a Newton-Raphson
method, using as initial guesses the soliton solutions computed in
Sec.~\ref{Sec:Solitary:Waves} or the analytical solution given in
\refeq{Eq:Esirkepov}. Our algorithm takes advantage of the sparsity
of the  Jacobian of $\bm{g}$, which was computed analytically, and
it carries out its LU factorization using parallel computation. In
our calculations we used $N_x = 1501$, $N_t = 150$ and a solution
$\bm{b}^*$ was accepted as valid if the tolerance of the Newton
method, i.e. the residual $\max(|\bm{g}(\bm{b}^*)|)$ of
\refeq{eq:algebr:quasisol}, was smaller than $5\times10^{-10}$.

\subsection{Quasi-soliton solutions}

We now provide examples of the use of the numerical algorithm of
\refsect{Sec:Quasi:Algorithm} in order to locate quasi-solitons.
Using the method introduced in Sec.~\ref{Sec:Solitary:Waves} we find
that, for $V = 0.9$, there is a soliton solution with $p=1$ at
$\omega = \omega^*\equiv  0.95805238$. We then use these
parameter values and $L=400$, $\Omega=\omega^*$ with the code
described in Sec.~\ref{Sec:Quasi:Algorithm} and compute a solution
using as initial guess the spatiotemporal profile found through the
integration of \refeqs{Eq:Circular:a}{Eq:Circular:phi} and the use
of \refeq{eq:Ansatz} and
\refneq{eq:fluid:vars}-\refneq{eq:fluid:vars:Final}. We then use the
velocity as continuation parameter to compute quasi-solitons with
$\Omega=\omega^*$. At each value of $V$, the code used as initial
guess the quasi-soliton solution found at the previous velocity
value.

\begin{figure}[h]
\noindent\includegraphics[width=20pc]{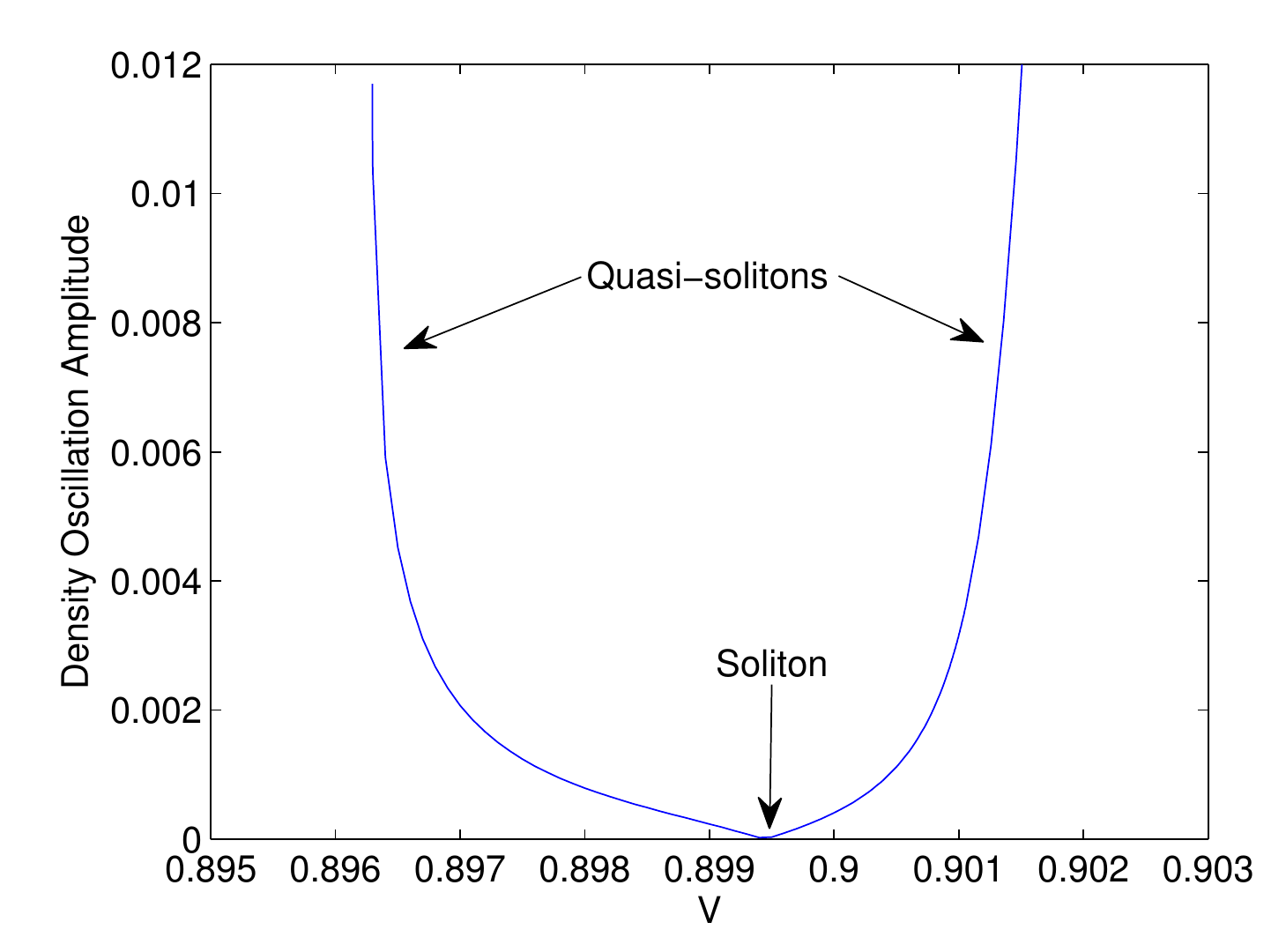}
\caption{Amplitude of the density oscillations at the quasi-soliton
tail versus $V$ for $\Omega=\omega^*$\label{Fig:Quasi:V} } .
\end{figure}

\begin{figure}[h]
\noindent\includegraphics[width=20pc]{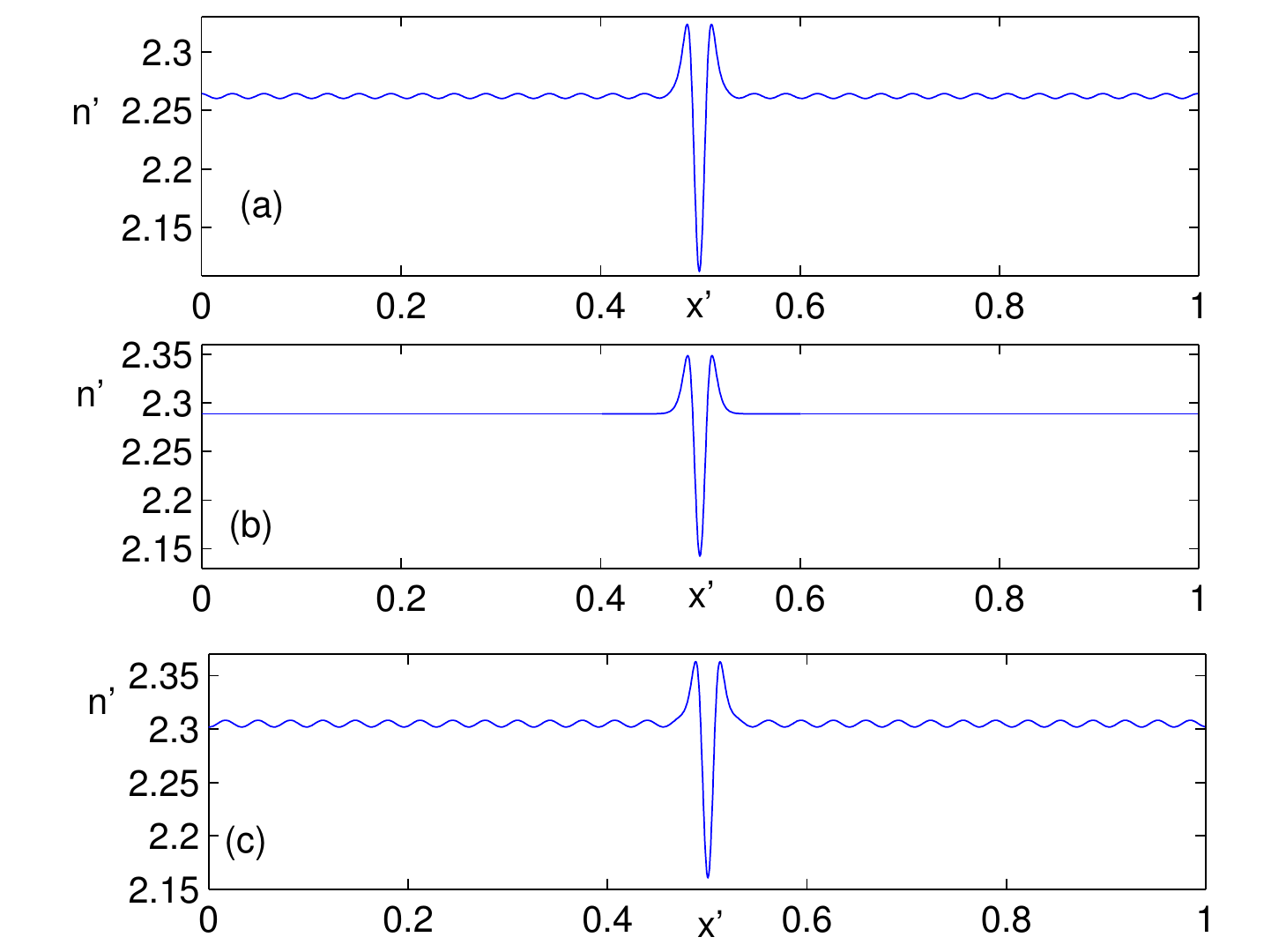}
\caption{Boosted frame density profile of $p = 1$ quasi-soliton
solutions at $t'=0$ with $\Omega = \omega^*$. Panels (a)-(c)
correspond to $V = 0.8970, 0.8995$ and $0.901$, respectively.
\label{Fig:Quasi:Density}}
\end{figure}

Figure \ref{Fig:Quasi:V} shows the relative (mean value was
subtracted) amplitude of the density oscillations at the
quasi-soliton tail versus the velocity at $\Omega=\omega^*$. The
amplitude of the density oscillations at the tail is equal to
$3.36\times 10^{-5}$ at $V = 0.8995$. The limiting factor in our
computations is the discretization error in space which is
$~\left(\Delta x\right)^2=0.07$. Therefore, this result is in
agreement with the solution of
\refeqs{Eq:Circular:a}{Eq:Circular:phi}, i.e. a fully localized
solution at $V=0.9$. Panels (a)-(c) in Fig.~\ref{Fig:Quasi:Density}
display the electron density profile of quasi-solitons with
$\Omega=\omega^*$ and velocity $V = 0.8970, 0.8995$  and $0.9010$,
respectively. The oscillations at the tail except at a particular
value of $V$ [panel (b)] are evident. The wavelengths of the
oscillations at the tail are in agreement with the linear analysis
of Sec.~\ref{Sec:Geometric:Arguments}. They are about $2\pi
V/\sqrt{1-V^2}$, i.e. they correspond to the eigenvalues
$\lambda_{3,4}$ that are associated with small amplitude
oscillations of $\phi$ according to \refeq{Eq:phi:linearized}. These
numerical calculations demonstrate that the branches of solitons are
parametrically embedded at the continuous spectrum of the
quasi-soliton.

\begin{figure}[h]
\noindent\includegraphics[width=20pc, clip=true]{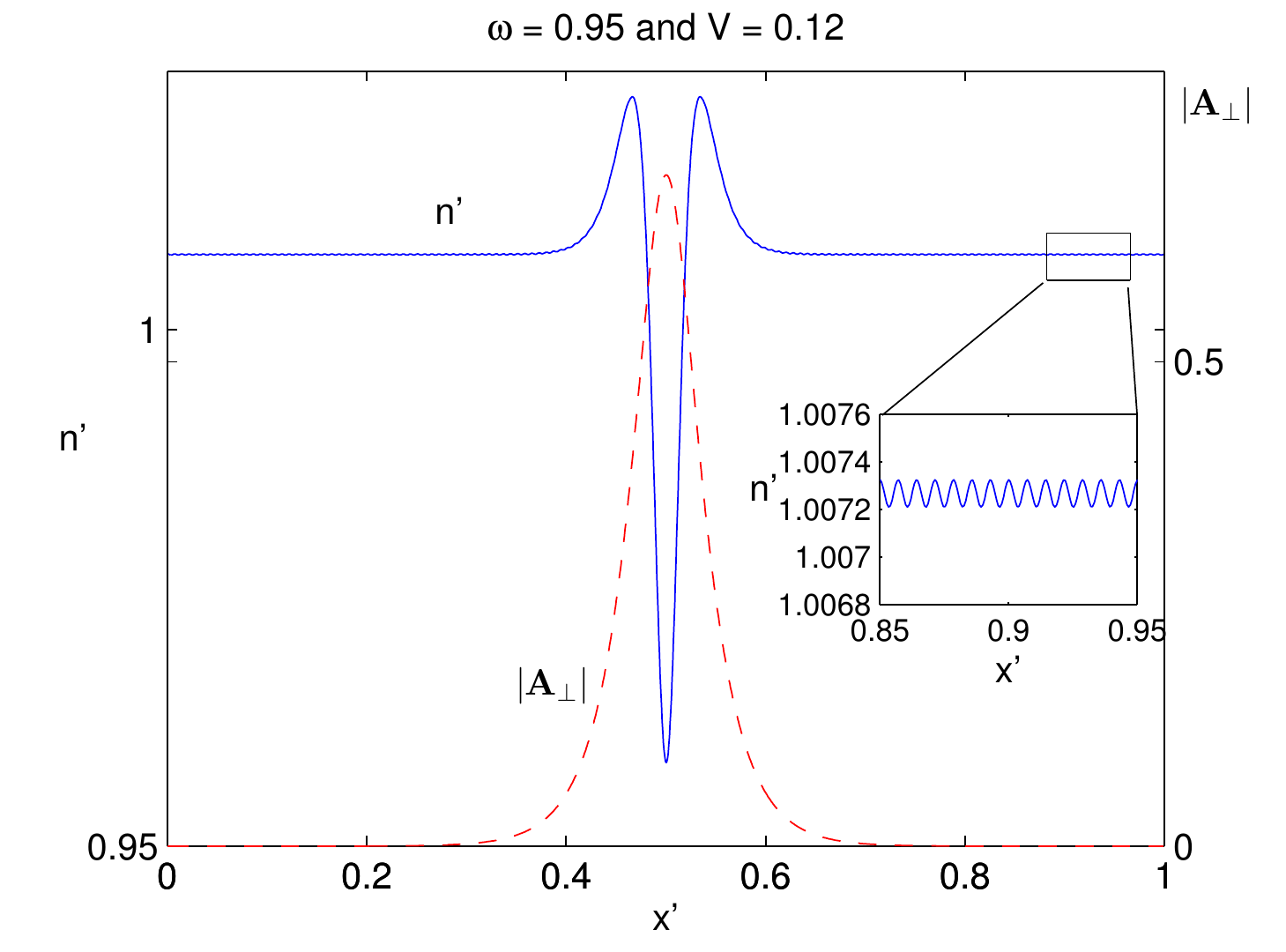} \caption{Density and
amplitude of the vector potential  profiles of a $p = 0$
quasi-soliton at $t'=0$. Parameter values are $\Omega = 0.95$ and
$V= 0.12$. \label{Fig:Quasi:P0}}
\end{figure}

The existence of $p=0$ quasi-solitons with finite velocity
was also investigated with the finite-difference code. For these
calculations we set $\omega=0.95$,  $L=100$ and used $V$ as a
continuation parameter. The analytical solution given by
\refeq{Eq:Esirkepov} was used as initial guess.
Solutions where found in a continuous range of $V$;
as an example, we show in Fig. \ref{Fig:Quasi:P0} a $p=0$ quasi-soliton with $V=0.12$,
thus giving numerical evidence of the existence of this special type
of quasi-soliton. The inset shows that small amplitude oscillations
with a high-wavevector exist at the tail. This result is again in
agreement with the linear analysis of Sec.
~\ref{Sec:Geometric:Arguments}. As $V\rightarrow 0 $ the wavelength
of the tail oscillation ($2\pi V/\sqrt{1-V^2}$) vanishes. As a
consequence, numerical difficulties arise at this particular limit,
which requires a high resolution to capture the structure of the
solution appropriately.

\subsection{Stability of the quasisoliton solutions\label{Sec:Quasi:Stability}}

In this section we study numerically
the stability of the quasi-soliton solutions.
Since these solutions are calculated with finite precision, the residual
in \refeq{eq:algebr:quasisol} acts as a perturbation of the exact solution.
Thus, integrating the Maxwell-fluid model \refeq{Sys:Fluid:0} with
quasi-soliton solutions determined using the algorithm of \refsect{Sec:Quasi:Algorithm}
as initial condition provides information on their stability.

The numerical integration is performed in the lab-frame with
the pseudo-spectral code described in \refref{siminos2014}.
For the spatial dependence of the field and plasma quantities, Fourier space
discretization is used, while time stepping is 
handled by an adaptive fourth order Runge-Kutta scheme.
In order to ensure numerical stability a filter in 
Fourier space of the form 
$\exp(-36\,(|k_x|L/(\pi\,N_x))^{24})$
is used. This prevents the growth of aliased high-$|k_x|$ modes,
while the dynamics of the physically relevant modes
are not affected. The solutions found in the boosted frame
with the algorithm of \refsect{Sec:Quasi:Algorithm} are transformed
to the lab frame using Lorentz transformations in order to obtain
the initial conditions required for the direct numerical integration.

\begin{figure*}[ht!]
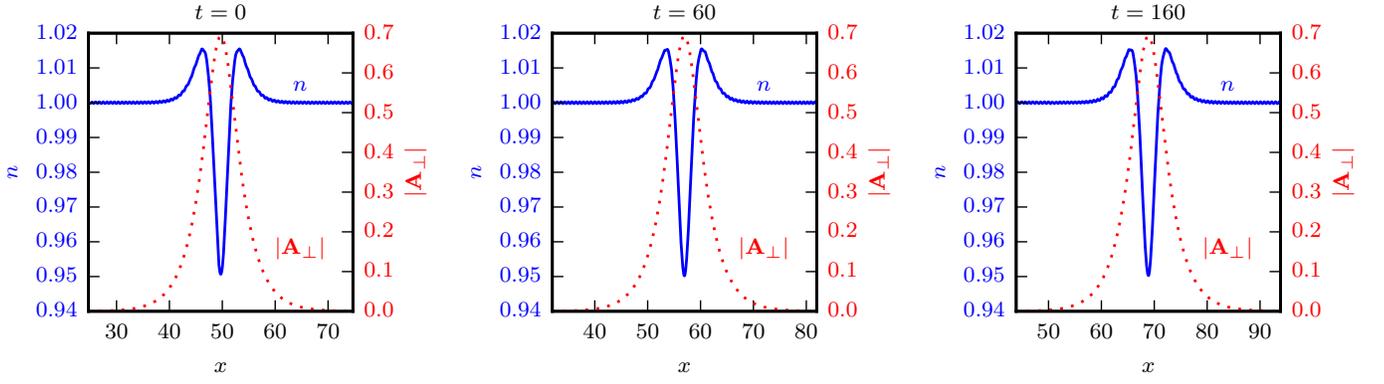

  \includegraphics[width=\textwidth]{{{qsoliton_P0_Omega_0.95_V_0.12_snapshots}}}
  \caption{\label{f:P0stab} Snapshots of propagation of $p=0$ quasi-soliton
  with $\Omega = 0.95$ and $V=0.12$.}
\end{figure*}

\begin{figure*}[ht!]
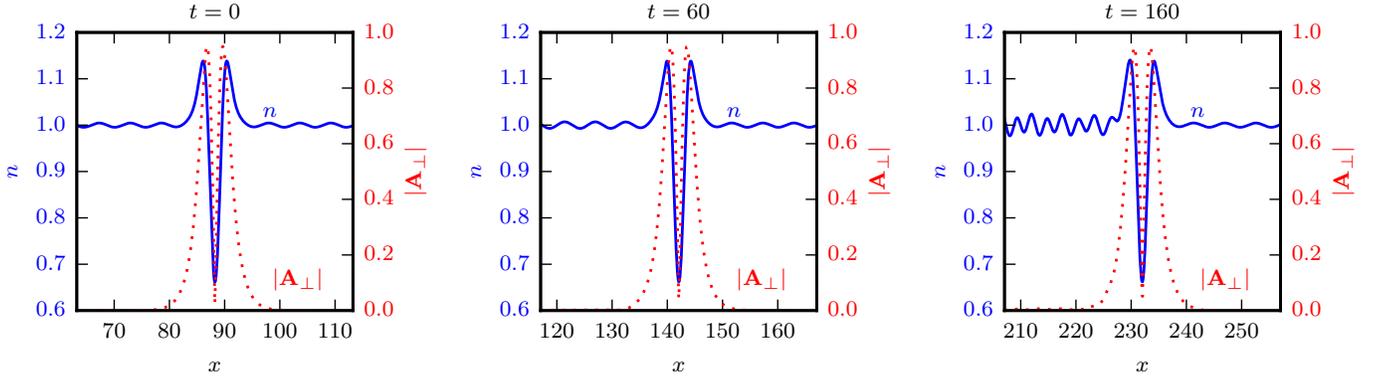

  \includegraphics[width=\textwidth]{{{qsoliton_P1_Omega_0.958_V_0.8970_snapshots}}}
  \caption{\label{f:P1stab}
  Snapshots of propagation of $p=1$ quasi-soliton with $\Omega=0.958$ and $V=0.897$.
  }
\end{figure*}

In \reffig{f:P0stab} we show snapshots from the evolution of a $p=0$
quasi-soliton with $\omega = 0.95$ and $V=0.12$. We observe that the
quasi-soliton remains essentially unchanged up to $t=160$,
indicating stability. 
In \reffig{f:P1stab} we show the evolution of a 
$p=1$ quasi-soliton with $\omega=0.958$ and $V=0.897$. 
An instability develops in the
trailing edge of the density profile. 
This instability is similar to the one
observed in fluid simulations of $p=1$ solitons~\cite{Saxena_06} and
connected to the forward stimulated Raman scattering
instability~\cite{Saxena_07}. The disturbance of the wake leads to
radiation of part of the electromagnetic field of the soliton.
We note that ion motion, not included here, is expected to
also play a role in quasi-soliton evolution, in particular for small $V$,
as already seen in studies of soliton stability at the ion time scale~\cite{lehmann2006}.
A more comprehensive study of the stability of quasi-solitons
is beyond the scope of this work and will be presented in a future
publication.

\section{Conclusions\label{Sec:Conclusions}}

We have shown that quasi-soliton solutions of the Maxwell-fluid
model exist in continuous domains of the $\omega\--V$ plane.
These solutions consist of a localized electromagnetic pulse with
$p=0,1,\ldots$ number of nodes of the vector potential
trapped in a plasma density cavity with oscillations at its tails.
Our study also sheds light to the organization in parameter space
of the circularly polarized relativistic solitons identified in previous studies.
The solitons with $p>0$ turn out to be special members of the corresponding
quasi-soliton family for which the density tail-oscillations vanish.
Therefore, such solitons exist for specific values  $V=V(\omega)$,
i.e. they form branches in the $\omega\--V$ plane. This is
consistent with the geometric arguments of the theory of dynamical systems
which were used in \refsect{Sec:Geometric:Arguments} to formulate a method to locate these branches.
Although the values of $\omega$ which give rise to localized solitons are
found numerically and with a finite accuracy, the bisection method developed
proves rigorously their existence because it is based on zero-crossings of
$a'(\xi_s)$ (even $p$) or $a(\xi_s)$ (odd $p$). This allows to rule out
the existence of a continuous spectrum for finite amplitude
and finite velocity, $p=0$ solitons. Such a continuous spectrum only
exists in either of the integrable limits of the Maxwell-fluid model,
i.e. for $\omega\rightarrow1$
or $V\rightarrow0$.
However, the only proper finite amplitude and velocity solitary waves with
$p=0$ are the quasi-solitons presented here.

Our stability study suggests that $p=0$ quasi-solitons are stable
while ones with $p>1$ are unstable. This could help explain the
abundance of single-humbed soliton-like structures in laser-plasma
interaction. It appears rather natural that during the process of
excitation of such structures, non-vanishing tail oscillations ('wakes') are
also excited by the driving laser pulse. We therefore expect that
our study will prompt the identification of quasi-solitons structures in
laser-plasma simulations and experiments.

\begin{acknowledgments}
This work was supported by the Ministerio de Econom\'ia y
Competitividad of Spain (Grant No RYC-2014-15357)
and by the Knut and Alice Wallenberg Foundation (\textsc{pliona} project).
\end{acknowledgments}

\appendix

\section{Solitons with mobile ions\label{Sec:Ion}}

\begin{figure}[h]
\noindent\includegraphics[width=20pc]{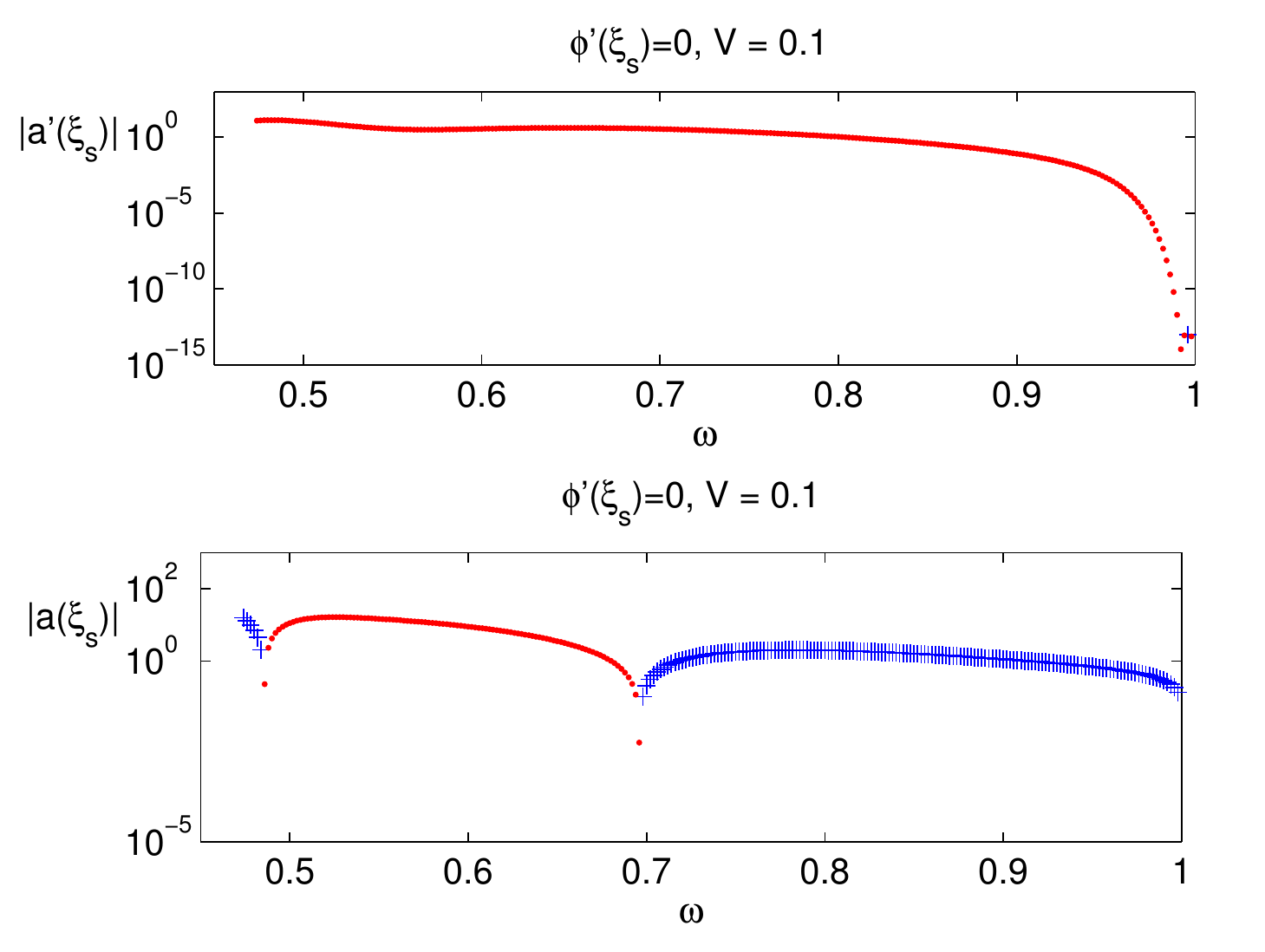}
\caption{Absolute value of $a'(\xi_s)$ (top panel) and $a(\xi_s)$ (bottom panel) versus
$\omega$ for $V=0.1$ (mobile ions)\label{Fig:Ions}}
\end{figure}

For mobile ions  Eqs.~(\ref{Eq:Circular:a})-(\ref{Eq:Circular:phi}) read
\cite{Kozlov_79}
\begin{subequations}
\begin{equation}
a''=\left[V\left(\frac{1}{R_e}+\frac{\rho}{R_i}\right)-\omega^2\right]a\label{Eq:Circular:a:Ion}
\end{equation}
\begin{equation}
\phi''=V\left(\frac{\psi_e}{R_e}-\frac{\psi_i}{R_i}\right)\label{Eq:Circular:phi:Ion}
\end{equation}
\end{subequations}
where $R_i = \sqrt{\psi_i^2-(1-V^2)(1+\rho^2a^2)}$, $\psi_i=1-\rho
\phi$ and $\rho = m_e/m_i$. In our calculations we will take
$\rho=1/1836$. The system admits the Hamiltonian
\begin{align}
H(a,p_a,\phi,p_\phi)=&
\frac{1-V^2}{2}\left[\left(\frac{p_a}{1-V^2}\right)^2+\omega^2a^2\right]\\
&-\frac{1}{2}p_\phi^2+V\left( R_e(a,\phi)+\frac{R_i}{\rho}\right)
\end{align}
and the involutions \refeq{Eq:Involution:1} and \refeq{Eq:Involution:2}.
Since the equilibrium $Q_0=(a=0,a'=0,\phi=0,\phi'=0)$ has
eigenvalues $\lambda_{1,2}=\sqrt{1+\rho-\omega^2}$ and
$\lambda_{3,4} = \sqrt{-(1-V^2)(1+\rho)/V^2}$, solitons are expected
to be organized in branches in the $\omega-V$ plane. The same
numerical method used for fixed ions was used to investigate the
spectrum of the solitons with mobile ions. The main difference is
the initial condition that now reads
\begin{equation}
\bm{x}_s(\xi=0)=\frac{\epsilon}{\sqrt{2-\omega^2+\rho}}\left[1 \ \
\sqrt{1-\omega^2+ \rho}\ \ 0 \ \ 0 \right]\label{Eq:IC:Mobile}
\end{equation}
Figure \ref{Fig:Ions} shows the Poincare maps ($\phi'(\xi_s)=0$) of
$a'(\xi_s)$ (top panel) and $a(\xi_s)$ (bottom panel ) versus
$\omega$ for $V = 0.25$. Since the variable $a'(\xi_s)$ does not
vanish, we conclude that $p=0$  solitons do not exist. The two zeros
of $a'(\xi_s)$ correspond to the  $p=1$ branch of solutions that has a
turning point in the $\omega-V$ plane (see Ref. \cite{Farina01a}).

\end{document}